\documentclass[10pt,twocolumn,letterpaper]{article}

\usepackage{wacv}              % To produce the CAMERA-READY version
\usepackage{multirow} % For \multirow
\usepackage{amssymb} 
\usepackage[table,xcdraw]{xcolor}  % 加 table 支持表格着色
\usepackage{makecell} 
\usepackage{graphicx}
\usepackage{amsmath}
\usepackage{amssymb}
\usepackage{threeparttable}
\usepackage{booktabs}
\usepackage[skip=3pt]{caption}
\usepackage{svg}
\usepackage[inline]{enumitem}
\setlist{topsep=0pt, leftmargin=*, noitemsep,parsep=0pt,partopsep=0pt}
\usepackage[pagebackref,breaklinks,colorlinks]{hyperref}

\usepackage[capitalize]{cleveref}
\crefname{section}{Sec.}{Secs.}
\Crefname{section}{Section}{Sections}
\Crefname{table}{Table}{Tables}
\crefname{table}{Tab.}{Tabs.}

\def\wacvPaperID{88} % *** Enter the WACV Paper ID here
\def\confName{WACV}
\def\confYear{2026}

\title{Ultra-lightweight camera for real-time imaging in-the-wild}
\title{Enabling Practical Ultra-thin Metalens Imaging through Burst Fusion and Multi-Frame Restoration}
\title{Enabling Practical In-the-Wild Imaging with Ultra-Compact Metalens Cameras via Burst Fusion and Restoration}
\title{Enabling High-Quality In-the-Wild Imaging from Severely Aberrated Metalens Bursts}

\author{Debabrata Mandal \quad Zhihan Peng \quad Yujie Wang \quad Praneeth Chakravarthula \\
UNC Chapel Hill\\
{\tt\small \{debman, zp59, wyujie, praneeth\}@unc.edu}
}

\begin{document}
\twocolumn[{
\renewcommand\twocolumn[1][]{#1}

\maketitle

\begin{center}
    \vspace{-3mm}
    \setlength{\abovecaptionskip}{4pt}
    \centering
    \includegraphics[width=\textwidth]{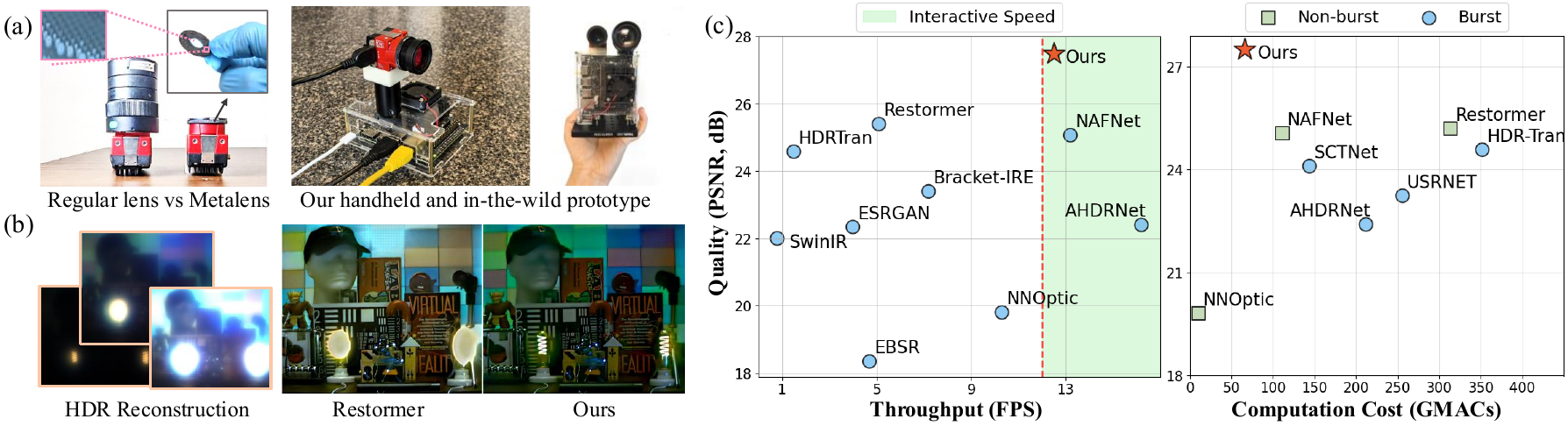}
    \captionof{figure}{\textit{Compact Metalens Camera with Real-Time Edge Performance.}
    \textbf{(a)} We jointly build a metalens, orders of magnitude thinner than a conventional camera lens, and a real-time burst image restoration method. Our camera module runs inference on a Jetson Nano Orin edge device.
    % Comparing (left) against a refractive convex lens our metalens is orders of magnitude thinner; We mount the camera module on a Jetson Nano Orin edge device for inference. 
    \textbf{(b)} We demonstrate in-the-wild imaging, including HDR reconstruction, as an avenue for our burst restoration pipeline. 
    \textbf{(c)} Our real-time image restoration on the handheld metalens camera outperforms prior state-of-the-art methods.
    }
    \label{fig:teaser}
    \vspace{1mm}
\end{center}
}]


\maketitle

\def\modelName{\textsc{None}}
\def\dataName{\textsc{MetaHDR}}

%%%%%%%%% ABSTRACT
% \vspace{-3mm}
\begin{abstract} 
\vspace{-4mm}

We tackle the challenge of robust, in-the-wild imaging using ultra-thin nanophotonic metalens cameras. 
% As the need for ever-smaller imaging systems intensifies, metalenses---planar optics composed of arrays of nanoscale scatterers---have recently emerged as a compelling alternative to conventional bulky refractive elements.
Meta-lenses, composed of planar arrays of nanoscale scatterers, promise dramatic reductions in size and weight compared to conventional refractive optics.
% However, they are far from practical adoption due to severe chromatic aberrations, unwanted light scattering, limited spectral bandwidth, and low light throughput and efficiency.
However, severe chromatic aberration, pronounced light scattering, narrow spectral bandwidth, and low light efficiency continue to limit their practical adoption.
In this work, 
% we draw inspiration from the ubiquitous burst-capture pipelines in modern smartphones to overcome these limitations.
we present an end-to-end solution for in-the-wild imaging that pairs a metalens over $12000\times$ thinner than conventional optics with a bespoke multi-image restoration framework optimized for practical metalens cameras.
% \todo{is it 10K times thinner? or 50K? can you really compute the size difference for your exeprimental lenses?. Regular lenses are around 6.3 mm and our metalens would be around 350 nm} 
Our method centers on a lightweight convolutional network paired with a memory‐efficient burst fusion algorithm that adaptively corrects noise, saturation clipping, and lens-induced distortions across rapid sequences of extremely degraded metalens captures.
Extensive experiments on diverse, real-world handheld captures demonstrate that our approach consistently outperforms existing burst-mode and single-image restoration techniques. 
These results point toward a practical route for deploying metalens‐based cameras in everyday imaging applications.
% These results pave the way for practical deployment of metalens cameras in everyday imaging applications.
Project page: \href{https://codejaeger.github.io/metahdr}{codejaeger.github.io/metahdr}.

% Metalenses offer nanoscale control of light, enabling ultra-thin, lightweight optics that could revolutionize handheld consumer imaging, and augmented and virtual reality. 
% However, their adoption is hindered by severe chromatic aberrations, light scattering, limited broadband performance, and low optical efficiency. 
% Burst imaging, widely used in smartphone cameras, enhances handheld photography by reducing noise, improving dynamic range, and increasing resolution. 
% Building on these insights, we design and prototype a $10,000\times$ thinner metalens compared to conventional compound optical lenses and introduce a multi-image restoration framework for noise, saturation and aberration correction, specifically tailored for handheld metalens cameras. 
% Our framework features lightweight networks, memory efficient burst fusion and adaptive correction techniques to restore high-quality images from extreme degraded metalens bursts.
% We benchmark our framework on a real in-the-wild collected dataset and show superior performance over existing burst or image restoration techniques. 
\end{abstract}

%%%%%%%%% BODY TEXT
\section{Introduction}
\label{sec:intro}

\vspace{-1.5mm}

Cameras have evolved from bulky mechanical systems with multi-element assemblies to today's slim, high-performance mobile systems, enabled by advances in stacked optics, high-resolution sensors, and optical stabilization.
Further miniaturization could enable seamless integration into wearables, smartphones, drones, and IoT (internet-of-things) devices, allowing always-on, context-aware sensing without bulky form factors. 
Unfortunately, achieving aberration-free, high-quality images still depends on complex multi-element lens stacks, creating a fundamental barrier to size reduction.
Metalenses, which are planar arrays of nanoscale scatterers that shape optical wavefronts at subwavelength scales, offer a promising path toward \textit{ultra-compact, flat optics} (see \cref{fig:teaser}).
However, their practical adoption is hindered by intrinsic hyperchromaticity, which causes severe chromatic aberration \cite{presutti2020focusing}, and by low optical efficiency arising from scattering losses and fabrication imperfections \cite{xu2022influencing}.

% Camera technology has evolved from bulky mechanical systems to compact, high-performance mobile devices with professional-grade imaging capabilities through advances in stacked optics, high-resolution sensors, and optical stabilization. 
% However, achieving aberration-free, high-quality images still requires complex multi-lens systems that limit further miniaturization—a critical barrier for emerging applications in wearables, AR/VR, and next-generation mobile imaging. 
% Novel optical solutions that transcend traditional lens limitations are needed to achieve superior image quality in ultra-compact form factors. Metalenses provide a promising alternative to traditional bulky lens stacks, leveraging nanoscale wavefront manipulation to enable ultra-compact and cost-effective camera designs, see \Cref{fig:teaser}. Despite their potential, imaging performance is hindered by significant chromatic aberrations due to inherent hyperchromaticity \cite{presutti2020focusing}, as well as reduced optical efficiency from scattering and fabrication imperfections \cite{xu2022influencing}. 

Recent advances in metalens design have greatly expanded their functionality, enabling broadband imaging \cite{seo2024deep,froch2024beating,dong2024achromatic}, wide field-of-view capture \cite{fullwfov}, extended depth-of-focus optics \cite{bayati2022inverse}, light field imaging \cite{lin2019achromatic}, and on-sensor integration \cite{chakravarthula2023thin}. 
However, these systems rely on \textit{computationally intensive reconstruction pipelines}, 
% making them impractical for \textit{resource-constrained edge devices}, such as smartphones \cite{froch2024beating,chakravarthula2023thin}.
% As a result, most approaches \cite{tseng2021neural, zhang2024deep} are developed and evaluated under \textit{controlled, low-dynamic-range (LDR) settings}, typically imaging display screens and relying on heavy post-processing. 
% Such pipelines
and
struggle to generalize to \textit{unconstrained environments} with extreme lighting variations, from bright outdoor scenes to dim indoor settings. 
% For instance, as shown in \Cref{fig:cam2}.b (top), an indoor OLED testbed exhibits a limited luminance range that fails to capture the wide dynamic range of natural scenes, leading to significant performance degradation when models are deployed in the wild.
% For instance, as shown in \Cref{fig:cam2}.b (top), an indoor OLED testbed's narrow luminance range fails to represent the full dynamic range of natural scenes, resulting in severe performance degradation when these models are deployed in the wild. \todo{not sure if fig 2 is adding a lot here. fig 2 also is incomplete. What does count mean? the illustration and labels do not correctly communicate the dynamic range issues. You can put it in suppl. and refer imo}
% 
% Compounding this, supervised training of image restoration networks requires large-scale paired captures from both metalens and high-quality compound-optic cameras (to establish ground truth).
% Collecting such datasets outdoors is particularly challenging: varying illumination and dynamic range complicate exposure matching, while parallax and depth-induced motion introduce misalignments and shifting homographies across frames.
% 
Compounding this, training these networks demands paired metalens-compound optic captures, which is particularly challenging in outdoor settings due to high dynamic range (HDR), exposure mismatches, lighting variability, and depth-induced parallax causing shifting homographies across captures.
% Collecting such data at scale is particularly challenging in outdoor settings, where varying illumination, dynamic range, and scene geometry make exposure matching difficult.
% Parallax and depth-induced motion blur between captures further introduce spatial misalignments and shifting scene homographies across captures, complicating direct supervision. 
On the other hand, although recent deep learning methods for HDR address brightness extremes \cite{perez2021ntire, perez2022ntire}, they are either:
\begin{enumerate*}[label=(\roman*)]
\item \textit{unscalable} to broader degradations beyond dynamic range issues, or
\item \textit{opaque black-box models} with limited interpretability and high computational cost \cite{zhang2024exposure, liu2023joint, lecouat2022high}.
\end{enumerate*}
As a result, existing pipelines remain ill-suited for general-purpose metalens imaging in-the-wild.

% In this work, we seek to overcome these challenges by revisiting modern imaging pipelines on resource-constrained edge devices. 
% Drawing inspiration from the widely adopted burst image processing on smartphones, we develop an end-to-end imaging pipeline for high-quality metalens imaging.
% While multi exposure burst captures overcome the low light efficiency and dynamic range issues of metalenses, they come with extreme degradations from noise, scattering and chromatic aberrations. To this end, we design a metalens to optimize the focusing across visible wavelengths, and design robust multi-exposure burst fusion pipeline for dealing with extreme degradations. 
% We circumvent the necessity of paired data captures for supervised learning by making our pipeline rely only data-agnostic features for image processing. 
% Moreover, we decouple the job of restoration and HDR fusion into separate modules connected by a pixel correction unit relying on softmax-based confidence weighting. This lets us keep each module within bare minimum compute requirements leading to better and accessible model design blueprints.
% Overall, we showcase a practical camera prototype featuring a single metalens that can be manufactured easily and assembled in-house. 

In this work, we address the core challenges of metalens imaging by introducing a compact, end-to-end pipeline for high-quality, in-the-wild capture.
Inspired by burst photography in smartphones, we use \textit{multi-exposure burst captures} to overcome the low light efficiency and narrow dynamic range of metalenses. 
% Drawing inspiration from widespread success of burst photography in smartphones, we leverage \textit{multi-exposure burst captures} to address the inherently low light efficiency and limited dynamic range of metalenses. 
Unlike conventional lenses, metalens bursts exhibit compounded degradations, including shot noise, chromatic aberrations, and subwavelength scattering, making standard burst fusion techniques ineffective.
To overcome these issues, we design and fabricate a metalens optimized for \textit{achromatic focus across the visible spectrum}, and jointly build a multi-stage computational pipeline tailored for \textit{burst fusion under extreme degradations}.

We decouple restoration and HDR fusion into lightweight, interpretable modules connected via a softmax-weighted pixel correction layer. 
This modular design delivers high-quality reconstructions with minimal computational overhead. 
Notably, this approach does not require paired metalens-compound optic training data.
% , thanks to \textit{domain-agnostic priors} and \textit{task-specific architectural biases}.
% \todo{due to... improve the argument if not correct.}
We validate our approach on a custom handheld prototype, featuring a single transmissive metalens, across diverse real-world scenes.
Our method consistently and significantly outperforms state-of-the-art burst and single-image restoration techniques, demonstrating the practical viability of ultra-thin metalens cameras for everyday imaging.

Our main contributions are as follows:
\begin{itemize}
    \item We introduce an efficient multi-exposure fusion framework tailored to metalens cameras, correcting noise, chromatic aberrations, and limited dynamic range with minimal compute overhead.
    % enabling practical deployment on thin resource-constrained systems.
    \item We demonstrate a fully functional, broadband ultra-thin nanophotonic camera built around a single tranmissive metalens, capable of high-quality imaging across diverse real-world conditions.
    \item We evaluate our approach against existing burst and single-image restoration methods, demonstrating superior performance in both simulated and in-the-wild captures.
\end{itemize}

\section{Related Work}
\label{sec:prior}

\paragraph{Flat-Optical Computational Cameras.} 
Optical miniaturization has revolutionized microscopy \cite{aharoni2019all}, spectroscopy \cite{yang2021miniaturization}, and photography \cite{galstian2014smart}, with today's smartphone cameras rivaling DSLRs in quality. 
However, traditional refractive systems require multiple lens elements for aberration correction, preventing further size reduction \cite{volkel2003miniaturized}.
% optics faces limitations as aberration correction requires multiple lens elements, increasing system bulk. 
Early lensless approaches replaced bulky optics with coded masks and computational reconstruction \cite{asif2016flatcam, boominathan2020phlatcam, antipa2017diffusercam}, but inherently lacked true focusing capability \cite{jain2025flattrack, shi2022seeing}. 
Metalenses have since emerged as a compelling solution, achieving high numerical apertures ($>$ 0.9) \cite{engelberg2020advantages} and enhanced signal-to-noise ratios \cite{chakravarthula2023thin,tseng2021neural,froch2024beating}
with a broad range of imaging and display applications
% This recent progress has generated a range of metalens-based imaging studies, from 
% end-to-end achromatic designs \cite{tseng2021neural} and wide field-of-view systems with diffusion restoration \cite{chakravarthula2023thin} to broadband \cite{froch2024beating} imaging and display applications 
\cite{tseng2021neural, seo2024deep, dong2024achromatic, zhang2024deep, qi2022all, chen2018broadband, park2024all, chakravarthula2023thin,gopakumar2024full}.
% However, there hasn't been any practically viable way of demonstrating HDR imaging for metalens cameras which exhibit extreme degradation and loss of resolution. Here, we present a mass-producible 1 cm diameter metalens camera, bridging nanophotonic design with practical imaging.
Despite these advances, extreme chromatic aberration and resolution loss still prevent their practical adoption. 
Here, we present a metalens camera paired with a multi-image restoration pipeline that bridges nanophotonic design with real-world imaging needs, enabling in-the-wild capture in an ultra-compact form factor.

\vspace{-10pt}

\paragraph{Joint Aberration Removal and HDR.}
Deep learning 
% approaches for HDR imaging have demonstrated impressive performance in both
has advanced HDR imaging in both
single-exposure~\cite{chen2021hdrunet, yan2019attention} and multi-exposure settings~\cite{yan2019attention, liu2021adnet, niu2021hdr,perez2021ntire, perez2022ntire},
% While single-exposure methods simplify data acquisition, multi-exposure techniques generally achieve high-quality results \cite{perez2021ntire, perez2022ntire}, despite challenges such as alignment errors and ghosting artifacts in bracketed fusion~\cite{gupta2013fibonacci}. 
with industry solutions like Google's HDR+ \cite{hasinoff2016burst} and Sony IMX490's on-sensor HDR processing demonstrating burst-based imaging solutions.
% underscore the practical value of burst-based imaging. 
Recent methods combine HDR with tasks such as denoising~\cite{liu2023joint}, super-resolution~\cite{tan2021deep}, and comprehensive restoration~\cite{zhang2024exposure}, and have extended HDR to specialized hardware such as event cameras~\cite{messikommer2022multi} and diffractive optics~\cite{sun2020learning}. 
However, most existing methods operate on images already processed by fixed image signal processors (ISPs) or ISO settings, which limits their generalization to novel lighting and downstream tasks~\cite{zou2023rawhdr,onzon2021neural}. 
In contrast, we introduce a bracketed burst fusion algorithm that
\begin{enumerate*}[label=(\roman*)]
\item functions independently of camera-specific settings,
\item jointly corrects metalens-induced aberrations, and
\item delivers high-fidelity HDR reconstructions.
\end{enumerate*}

\paragraph{Burst Matching.}
\vspace{-10pt}

Recent computational photography techniques have greatly enhanced burst imaging in consumer devices. 
Google's HDR+ pipeline~\cite{hasinoff2016burst} and its successors \cite{Monod_2021} enhance low-light performance by aligning and merging multiple frames. 
Multi-scale pyramid schemes~\cite{ipol.2023.460} achieve sub-pixel registration precision using Lucas-Kanade alignment for super-resolution. 
End-to-end approaches such as deep burst super-resolution~\cite{bhat2021deepburstsuperresolution} unify alignment, denoising, and upsampling in a single network for efficiency. 
Recent optical flow methods such as RAFT~\cite{teed2020raftrecurrentallpairsfield} and pyramid CNNs~\cite{ranjan2016opticalflowestimationusing, sun2018pwcnetcnnsopticalflow} handle complex scene motion, while detector-free matching with LoFTR~\cite{sun2021loftrdetectorfreelocalfeature} uses transformer-based attention to align frames in dynamic or low-texture scenes.
Building on these advances, we introduce a burst-matching framework tailored to metalens imaging, explicitly compensating for their hyperchromatic aberrations and distortions from subwavelength scattering to deliver robust, high-fidelity fusion.

\section{Compact Metalens Camera Prototype}
\label{sec:overview}
We develop and fabricate an ultra-compact metalens and integrate into a handheld camera prototype (see \cref{fig:teaser}). 
Below, we summarize the prototype design and our image formation model; full implementation details can be found in the Supplementary Material.
% \todo{make sure additional details are in suppl.}
% We first provide an overview of the fabricated metalens, hand-held prototype camera and the image formation model.

% \input{figures/cam}

\subsection{Metalens Design and Fabrication} 
\begin{figure}[t]
    \centering
    \includegraphics[width=\linewidth]{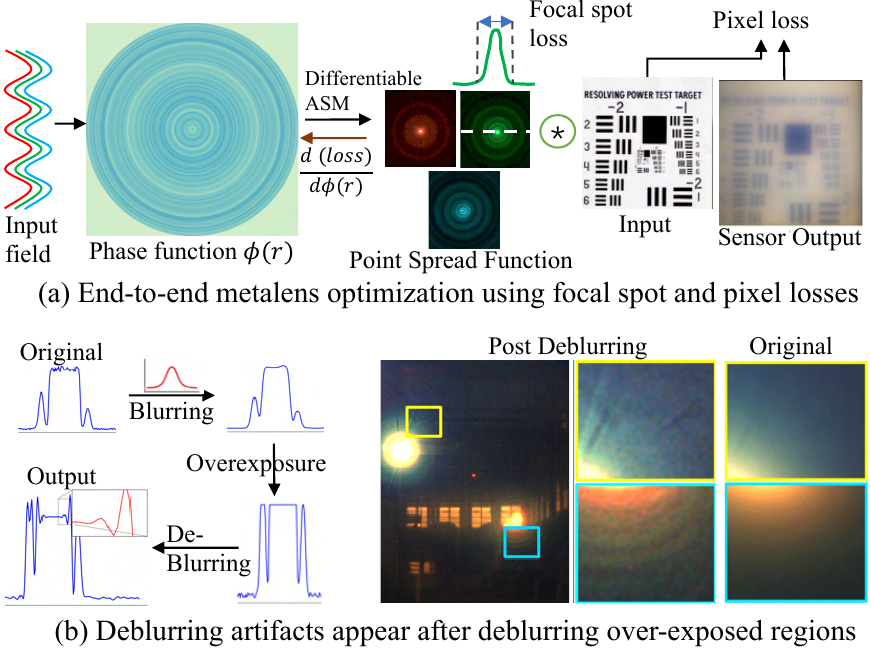}
    \caption{\textit{Metalens Optimization and Deblurring Artifacts.} 
    \textbf{(a)} We optimize the metalens phase profile via a radial parameterization, differentiable wave propagation, and joint focal spot (optical) and pixel-space (L2) losses.
    % Our metalens optimization involves a radially parametric phase design, a differentiable angular spectrum method propagation function and optical (focal radius) and pixel space (l2) losses. 
    \textbf{(b)} Naive deconvolution around saturated areas produces ringing and halo artifacts.
    % Deblurring artifacts arise due to naive deconvolution around saturated regions.
    }
    \label{fig:cam}
    \vspace{-2mm}
\end{figure}

We use a radially symmetric parameterization \cite{froch2024beating, chakravarthula2023thin} for the metalens phase $\phi(x_i, y_j) = \phi(r)$ for $i,j \in \{1,2,...,N\}$, 
% We parameterize the metalens using a simple radially symmetric per-pixel basis function similar to \cite{froch2024beating, chakravarthula2023thin}:
% \begin{equation}
%     \phi(x_i, y_j) = \phi(r), \quad r = \sqrt{x_i^2 + y_j^2}, \quad i,j \in \{1,2,...,N\},
% \end{equation}
where, $\phi(r)$ prescribes the local wavefront modulation at a  distance $r$ from the center and $N$ is the metalens discretization resolution. 
We optimize $\phi(r)$ for a 1cm aperture metalens via differentiable wave-propagation to minimize the focal-spot diameter
% This phase function is optimized following differential wave propagation principles \cite{} to reduce the lens focal spot size as shown in 
(see \cref{fig:cam}(a)). 
The optimized metalens is fabricated in-house
% Once optimized, we fabricated the 1 cm large-aperture meta-optic 
using standard nanofabrication techniques, see Supplementary Material.

% First, a silicon nitride layer is deposited by plasma-enhanced chemical vapor deposition (PECVD), then a high-resolution electron beam lithography (EBL) with ZEP520A resist and conductive polymer overcoat defined the metasurface pattern, followed by metal liftoff and inductively coupled plasma (ICP) etching.
% The process involved PECVD deposition of silicon nitride on quartz, high-resolution electron beam lithography (EBL) with ZEP520A resist and conductive polymer coating, followed by metal lift-off and ICP etching. 
% The resulting transmissive metalens is integrated with an Allied Vision 1800 U-510 CMOS sensor and paired with a Jetson Nano Orin board for handheld operation, as shown in \cref{fig:teaser}(a).
% The fabricated metalens was then mounted upon the camera module for evaluation. We then attach the camera module to a Jetson Nano Orin for a handheld capture setup as shown in \cref{fig:teaser}.a.

% \subsection{Portable Camera prototype}
% We attach the camera module to a Jetson Nano Orin  as in \cref{fig:teaser}.a via Alvium flex adaptor cable to connect the CSI-2 port of Jetson board (host side) to Allied Vision Alvium cameras (camera side). Compared to prior PC (or laptop) based cameras showed in \cite{chakravarthula2023thin, tseng2021neural} our embedded camera setup runs at higher video throughput and can also access a low-end gpu to support deep learning based post-processing. Further details on the camera-metalens assembly has been provided in the supplementary.

\begin{figure*}[!t]
    \centering
    \includegraphics[width=\linewidth]{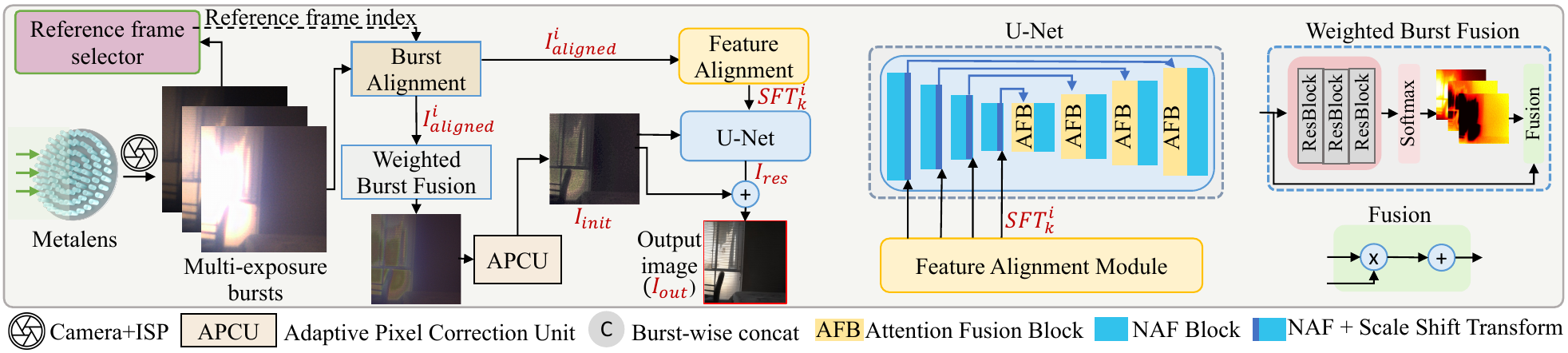}
    % \vspace{-4mm}
    \caption{\textit{Pipeline Overview.} Captured metalens bursts first go through a reference frame selector to identify the reference frame index ($r$). Each burst frame $I^i$ is then aligned to the reference, producing $I^i_\text{aligned}$. These aligned frames enter a weighted burst fusion module to produce a single fused image $I_\text{fused}$. An Adaptive Pixel Correction Unit (APCU) adjusts pixel intensities using a weighting operation, yielding $I_\text{init}$, which is finally refined by the restoration module to produce a high-quality output $I_\text{out}$. }
    \vspace{-3mm}
    \label{fig:pipeline}
\end{figure*}

\subsection{Image formation model} 
Metalens cameras suffer from severe chromatic and spatially varying aberrations, and a narrow dynamic range due to limited broadband efficiency and fabrication-induced scattering, all of which complicate image recovery.
Let $X(u)$ denote the true underlying scene radiance at pixel $u$. 
During the burst of $N$ captures, each frame $i$ undergoes motion-induced warp $W_i$ and exposure-dependent accumulation over time $\Delta t_i$.
We model the camera point spread function as $\mathbb{P}$ and additive sensor noise as $\eta_i$. 
After quantization to $q$ bits and per-frame ISP processing, the observed intensity is:
\begin{equation}
I_i(u) = \mathtt{ISP}\Bigl(\bigl\lfloor \Delta t_i,W_i\bigl(\mathbb{P} * X\bigr)(u) + \eta_i \bigr\rfloor_q\Bigr),
\end{equation}
where $\lfloor\cdot\rfloor_q$ clips intensity values to $[0,2^q-1]$. 
PSF-induced spatially varying blur and the clipping nonlinearity make naive deconvolution highly ill-posed: overexposed regions, once deblurred, often exhibit severe ringing and artifacts, see \cref{fig:cam}(b). 
Moreover, low light throughput \footnote{The percentage of transmitted to incident light energy} further narrows the measurable intensity and dynamic range, exacerbating quantization and clipping, and further degrading recovery fidelity.
Our reconstruction pipeline explicitly inverts these degradations across the burst to recover high-fidelity images.

\section{Burst Fusion and Image Recovery}
\label{sec:method}
\begin{figure}[t]
    \centering
    \includegraphics[width=\linewidth]{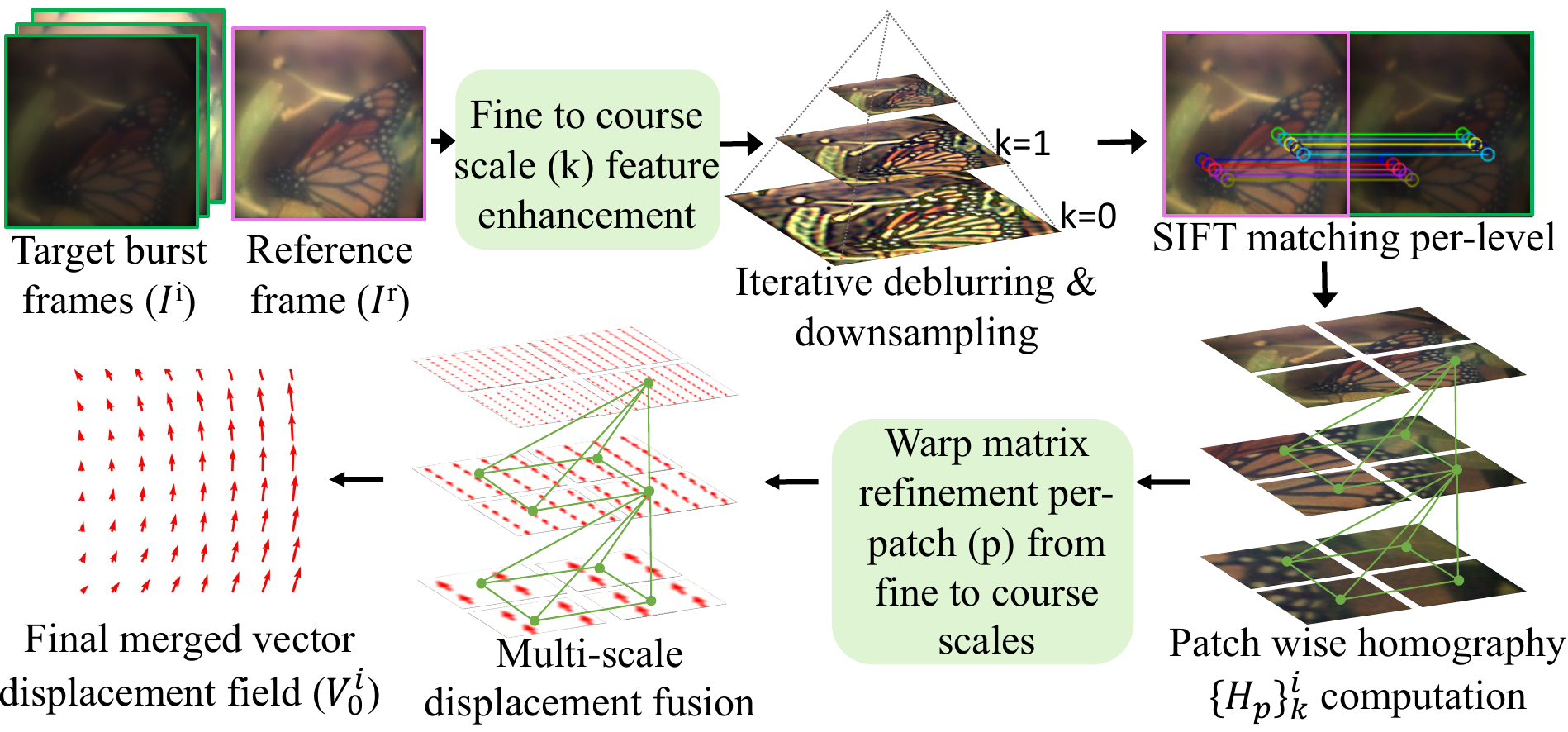}
    \caption{\textit{Burst Alignment.} 
    At each pyramid level, we first sharpen features via deconvolution before performing SIFT-based feature matching. 
    We then compute local patch homographies and fuse them across neighboring patches and scales to achieve precise frame registration.
    % We perform a per-scale feature enhancement using deconvolution to sharpen image features which are then used for SIFT based feature matching. The homography matrix computed per-patch from these matches is merged with neighbors and finer 
    }
    \label{fig:bmatch}
    \vspace{-2mm}
\end{figure}

\begin{figure}[h]
    \centering
    \includegraphics[width=\linewidth]{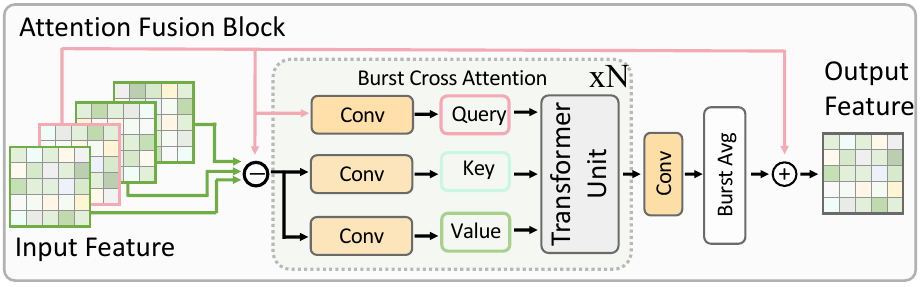}
    \caption{\textit{Burst Cross-Attention Fusion.} 
    We fuse aligned burst features via cross-attention, using the reference frame as query and others as key/value, through the skip connections.
    % Fusing the selective features for each burst frame coming from the  skip connections.
    }
    \vspace{-3mm}
    \label{fig:module_afb}
\end{figure}

Our burst image restoration pipeline is shown in \cref{fig:pipeline}.
First, we align a burst of images via patch-wise feature matching (see \cref{fig:bmatch}) and fuse the registered frames with a lightweight residual network to obtain an initial estimate. 
Next, a channel-efficient feature alignment module (\cref{fig:module_fab}) and a compact U-Net with attention fusion blocks (\cref{fig:module_afb}) correct residual aberrations and recover fine image details.

% \todo{It's better to give an overview of your overall pipeline before describing sub-modules. Briefly describe what are the key modules, what are their functions within the pipeline.}

% In this section, we give a brief overview of our burst alignment and restoration framework. \Cref{fig:pipeline} depicts the various parts of our burst pipeline.

\subsection{Burst Frame Alignment}

Effective burst fusion relies on robust frame alignment.
Our alignment pipeline---designed for \textit{efficiency}, \textit{simplicity}, and \textit{robustness}---operates on a multi-level image pyramid $\{I^i_k\}_{k=1}^K$ ($i$ being the frame index), where each level $k$ is iteratively deblurred and downsampled to expose stable features for reliable patch-wise matching (see \cref{fig:bmatch}). 
% Instead of naive downsampling, we iteratively downsample with deblurring yielding more image features up until the last level as shwon in \cref{fig:bmatch}. 

\noindent
\textbf{Reference frame selection:} The reference frame exposure time is first determined using the camera ISP's autoexposure algorithm. Given this reference exposure, we capture burst sequences at varying digital gains, similar to HDR+ \cite{hasinoff2016burst}. The reference frame is then selected as the sharpest frame from the burst using a gradient-based sharpness metric \cite{joshi2010seeing} applied to the green channel of the raw image data.

\noindent
\textbf{Iterative deblurring and downsampling:} 
We find that the standard SIFT feature matching algorithm \cite{lowe1999object} struggles on metalens images, where severe aberrations and blur obscure key points \cite{tang2014improving}.
To overcome this, we introduce a multi-scale deblurring and downsampling enhancement that reveals stable features at each pyramid level, dramatically increasing reliable matches across frames.
% feature enhancement technique to increase the overall feature matches across the image.
% \todo{Describe why you introduce this design, what do you want to deal with this design} 
At each level $k$, we perform Tikhonov-regularized deconvolution:
\begin{equation}
    \widetilde{I}^i_k = \min_{I^i_k} \left\lVert \widetilde{I}^i_{k-1}\downarrow_2 - \rho * I^i_k\right\rVert^2_2 + \lambda_k\left\lVert I^i_k \right\rVert^2_2,
    \label{eq:deconvolution}
\end{equation}
where $\rho$ is the PSF and $\downarrow_2$ denotes $2 \times$ downsampling. 
Solved in close form via fast Fourier transforms (FFTs), this step sharpens features at each scale, ensuring robust patch-wise matching, see Supplementary Material.

% The closed form solution with fast Fourier operations restores key features at each scale for reliable matching, see Supplementary Material. 
% \todo{where is closed form solution discussed? Refer to that in Suppl.}

\vspace{1mm}
\noindent
\textbf{Patch-wise homographies:} 
At each pyramid level, we compute local homographies $\{H_p\}_k^i$ over a grid of patches via feature matching, then derive per-patch displacement vectors $\{\vec{V}_p\}_k^i$ from these homographies. 

% \vspace{1mm}
% \noindent
% \textbf{Multi-scale displacement fusion:}
% To combine estimates across scales, we update each patch's displacement by blending fine- and coarse-scale cues:
% Given the downsampled and deconvolved images, we estimate patch-wise homographies $\{H_p\}_k^b$ at each scale. The patch-wise displacement vectors $\{\vec{V}_p\}_k^b$, derived from $\{H_p\}_k^b$, are aggregated across levels using:
% \begin{equation}
%     \label{eq:displacement_update}
%     % \vec{V_p}_k^b = \mathtt{best}(\vec{V_p}_k^b, \mathtt{neighbor_{p'}}\{\vec{V_{p'}}_k^b\}) + \omega \cdot \vec{V_p}_{k-1}^b,
%     \vec{V_p}_k^b = (1 -\omega_p)\,\mathtt{best}\bigl(\vec{V_p}_k^b,\mathtt{neighbor_{p'}}\{\vec{V_{p'}}_k^b\}\bigr) + \omega_p \,\vec{V_p}_{k+1}^b ,
% \end{equation}

\vspace{1mm}
\noindent
\textbf{Multi-scale displacement fusion:}
To integrate motion estimates across scales, we update each patch's displacement by blending coarse- ($\vec{V}_{p,\, k+1}^i$) and fine-scale ($\vec{V}_{p, \,k}^i$) cues:
% \begin{equation}
%     \label{eq:displacement_update}
%     % \begin{split}
%         \vec{V}_{p,\, k}^i = (1 -\omega_p)\,\mathtt{best}\bigl(\vec{V}_{p,\,k}^i, 
%         \:\:\:\: \{\vec{V}_{p',\,k}^i\}_{p' \in \mathtt{neighbor(p)}}\bigr) + \omega_p \,\vec{V}_{p,\,k+1}^i ,
%         % \nonumber
%     % \end{split}
% \end{equation}
\begin{equation}
    \label{eq:displacement_update}
    \vec{V}_{p,k}^i = (1-\omega_p)\,\mathtt{best}(\vec{V}_{p,k}^i, \{\vec{V}_{q,k}^i\}_{q \in \mathcal{N}(p)}) + \omega_p \,\vec{V}_{p,k+1}^i
\end{equation}
where $\omega_p$ balances the displacement refinement from finer scale (k) with that obtained from coarser scale (k+1) and $\mathtt{best}$ selects the optimal displacement from the patch (p) and it's neighbors denoted by $\mathcal{N}(p)$, through a shear ratio consistency test to discard outliers. The frame aligned to the reference frame  given by $I^i_\text{aligned}$ is obtained via pixel-wise warping with full-frame displacement map $V_0^i$ at the finest scale ($k=0$). By interleaving deblurring and downsampling, this approach uncovers reliable feature matches even at extreme exposures (\cref{fig:Align_arrows}(b)), far beyond what scale-invariant SIFT matching alone can achieve under severe metalens degradations.
A full algorithmic description is available in the Supplementary Material.

% Moreover, although SIFT is scale-invariant, our interleaved deblurring and downsampling reveal new matches at each lower scale, robustly for any exposure level, as shown in \Cref{fig:cam}.d. 

\subsection{Lightweight Real-time Burst Restoration}
% To achieve high-quality results on resource-constrained hardware, 
Our restoration pipeline employs a two-branch architecture (see \cref{fig:pipeline}).
The first branch generates an initial burst-fusion estimate $I_{\text{init}}$ via a compact residual network, while the second branch predicts a fine-detail correction $I_\text{res}$. 
The final restored image is obtained as $I_\text{out} = I_{\text{init}}+I_\text{res}$. 
% \todo{it's better to have an overview description about involved submodules that will be described below.}

% As shown in \cref{fig:cam}.c, a key point of difference in our framework against prior related works has been our focus on low-resource implementations so that the network can run provide real-time inference.

\vspace{1mm}
\noindent
\textbf{Initial burst fusion and adaptive pixel correction:}
Aligned burst frames $I^i_{\text{aligned}}$ (from previous module) are first combined by a lightweight residual network: 
% \todo{(term $I_{aligned}$ is never defined. Add an equation in the previous section defining it. codefined in 4.1)}
\begin{equation}
I_{\text{fused}} = \sum_i w^i \odot I^i_{\text{aligned}},
\label{eq:residual}
\end{equation}
% \vspace{-2mm}
where the fusion weights $w^i$ are produced by a series of residual blocks followed by a softmax normalization layer. 
An Adaptive Pixel Correction Unit (APCU) then rescales each pixel based on a learned confidence map $\in [0,1]$, producing the initial estimate $I_{\text{init}}$. Additional details are provided in the Supplementary Material.
% \todo{Could you describe APCU a little more? I saw that there are ablation results for this module, then it should be described more, include why you used it and its functions.}
% \textcolor{blue}{it's provided in the supplementary}

% As shown in \cref{fig:pipeline} the burst images after alignment, follow two separate branches.
% The first step involves getting an initial estimate (\textit{$I_{ini}$}) of the burst fusion using a lightweight residual network module and then restoring the fine-level details in the image using an image residual ($I_{res}$) as,
% \begin{equation}
%     I_{out} = I_{ini} + I_{res}
% \end{equation}
% We design our initial estimator module as a residual network module which takes as input the aligned burst frames $I^b_{aligned}$ and passes them through a series of residual blocks followed by a softmax weighting layer to generate the fusion weights ($w$) to get the fused image as,
% \begin{equation}
%     I_{fused} = I^b_{aligned} * w^b.
%     \label{eq:fusion}
% \end{equation}
% Next, we pass that through an adaptive pixel correction unit (APCU) which re-scales intensity values based on a confidence value \{$0..1$\} to give us the initial image estimate $I_{ini}$. 

 \begin{figure}[t]
    \centering
    \includegraphics[width=\linewidth]{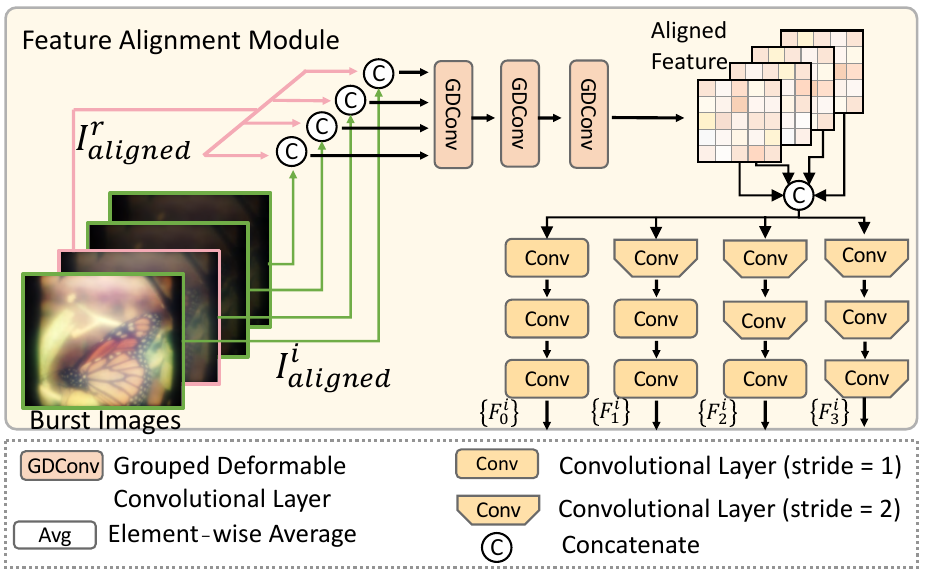}
    \caption{\textit{Selective Feature Alignment}. Our feature alignment module extracts and integrates burst-frame features conditioned on the reference frame to correct residual defects 
    in the initial estimate $I_\text{init}$.
    }
    \label{fig:module_fab}
    \vspace{-3mm}
\end{figure}

\vspace{1mm}
\noindent
\textbf{Conditioned U-Net residual refinement:}
The second branch refines $I_{\text{init}}$ with a compact U-Net built from lightweight NAF blocks \cite{chen2022simplebaselinesimagerestoration}.
Rather than encoding the entire burst, we concatenate the reference frame's displacement field $V^i$ (from \cref{eq:displacement_update}) with the aligned burst frames and pass these through a feature alignment module (see \cref{fig:module_fab}).
The U-Net then takes $I_{\text{init}}$ as input and predicts a residual correction $I_\text{res}$.
By operating solely on a single fused image, this conditioned U-Net significantly reduces memory consumption and improves runtime latency.
% \todo{you need to use these notations in the fig. otherwise it is very confusing.}

% The second branch involves a lightweight U-Net implemented using a variant of NAFNet \cite{chen2022simplebaselinesimagerestoration} which replaced heavy residual blocks using lightweight NAF blocks for image restoration applications. 
% The displacement fields $V^b$ computed from the burst alignment module are concatenated with the burst frames and passed into a feature alignment module as shown in \cref{fig:pipeline}. 
% At this point, the Unet takes in a single image i.e. $I_{ini}$ and predicts a residual correction to the initial image while being conditioned via the feature alignment module. 
% Taking this approach instead of feeding all burst frames through the U-net encoder saves reduces memory consumption and improves runtime latency significantly. 

\noindent
\textbf{Selective feature alignment and SFT fusion:}
To correct residual misalignments and enrich restoration, our feature alignment module (see \cref{fig:module_fab}) uses lightweight grouped deformable convolutions \cite{zha2023engd} to extract burst-frame features $F_k^i$ (from $i$-th frame) conditioned on the reference frame. 
At each U-Net encoder level $k$, we reduce $F_k^i$'s channels to one-fifth of the encoder's width, constraining the feature alignment module to a few select key features from each burst. We then inject them via Scale-Shift Feature Transform (SFT) \cite{wang2018recovering}: 
\begin{equation}
    SFT^i_k = \mathtt{scale}(F_k^i) \cdot e_k + \mathtt{shift}(F_k^i),
    \label{eq:sft}
\end{equation}
enabling spatially-varying modulation of the encoder features $e_k$.
This approach adaptively modulates the encoder activations to handle spatially varying aberrations and alignment errors that standard convolutions cannot.

\noindent
\textbf{Attention fusion via burst cross-attention:}
We propagate the transformed features through skip connections into an Attention Fusion Block (AFB) in the decoder. 
The AFB applies cross-attention along the channel dimension, using the reference frame as the query (Q) and the remaining burst frames as keys (K) and values (V), to efficiently merge multi-frame information.
% and then averages the frames (\cref{fig:module}). 
For an $H\times W \times C$ input, channel-wise attention operates in $O(HWC^2)$ time, dramatically reducing memory and compute compared to spatial attention's $O(H^2W^2C)$ cost \cite{zamir2022restormer}.
% 
% instead of performing attention over the spatial dimensions which would be $O(H^2W^2)$ for an input of size $H\times W \times C$, we perform it over the channel dimension (C) following \cite{zamir2022restormer} to keep memory consumption low. 
% Network implementation details are provided in the supplementary.

% 
This modular separation of fusion, pixel correction, and residual refinement yields high-quality reconstruction with minimal compute and memory overhead, making the pipeline ideal for resource-constrained edge devices.
% Comprehensive implementation details are provided in the Supplementary Material. 

 % \footnote{We avoid the \textit{$\mu$}-law tone-mapping loss proposed by \cite{kalantari2017deep} since it only produced marginal improvements over the unmapped image.}

\subsection{Training Objectives} 
We supervise both the initial estimate $I_\text{init}$ and the final output $I_\text{out}$ against the ground truth image $I_\text{gt}$ using an L1 loss:
\begin{equation}
    \mathcal{L}_\text{pixel} = \lvert| I_\text{out} - I_\text{gt} \rvert| + \lvert| I_\text{init} - I_\text{gt} \rvert|.
\end{equation}
Encouraging $I_\text{init}$ to match the ground truth provides intermediate supervision, which stabilizes training and allows the subsequent refinement network to remain lightweight.
To discourage the fusion network from relying on over-saturated pixels, we generate saturation masks $S^i$ (see \cref{fig:itw_adapt}) and penalize non-zero fusion weights in saturated regions:
\begin{equation}
    \mathcal{L}_\text{sat} = \sum_{k}\lvert| S^i. w^i \rvert|,
    \label{eq:resi_loss}
\end{equation}
where $w^i$ are the softmax fusion weights from \cref{eq:residual}. 
Finally, we include a perceptual loss $\mathcal{L}_\text{LPIPS}$ \cite{zhang2018unreasonable} to promote visual realism. The total loss is:
\begin{equation}
    \mathcal{L}_\text{total} = \mathcal{L}_\text{pixel} + \tau_1\mathcal{L}_\text{sat} + \tau_2\mathcal{L}_\text{LPIPS},
    \label{eq:loss}
\end{equation}
where $\tau_1$ and $\tau_2$ balance the saturation and perceptual terms.

% During training, both the final prediction $I_\text{out}$ and the output from the adaptive pixel correction unit $I_\text{init}$ is compared with the ground truth image ($I_\text{gt}$) using L1 loss:  
% \begin{equation}
%     \mathcal{L}_{pixel} = \lvert| I_\text{out} - I_\text{gt} \rvert| + \lvert| I_\text{init} - I_\text{gt} \rvert|.
% \end{equation}
% Furthermore, we construct saturation masks $S^i$ (as shown in \cref{fig:itw_adapt}) to penalize the weighed burst fusion module for assigning non-zero weights to over-saturated image regions, 
% \begin{equation}
%     \mathcal{L}_{residual} = \sum_{k}\lvert| S^i. w^i \rvert|
%     \label{eq:resi_loss}
% \end{equation}
% % \vspace{-2mm}
% where $w^i$ are the softmax weights from \cref{eq:residual}. 
% We also employ a perceptual loss \cite{zhang2018unreasonable}, contributing to the final weighted ($\tau$) loss in the optimization of parameters $\theta$.
% We also incorporate a VGG19-based perceptual loss \cite{zhang2018unreasonable}, optimizing network parameters $\theta$ through a final weighted ($\tau$) loss objective,
% \begin{equation}
%     \text{arg}\,\max\limits_{\theta} \, \mathcal{L}_{pixel} + \tau_1\mathcal{L}_{residual} + \tau_2\mathcal{L}_{lpips}.
%     \label{eq:loss}
% \end{equation}
% \vspace{-2mm}

\subsection{Robust Adaptation to In-the-Wild Conditions} 
\label{sec:generalize_itw}

\begin{figure}[t]
    \centering
    \includegraphics[width=\linewidth]{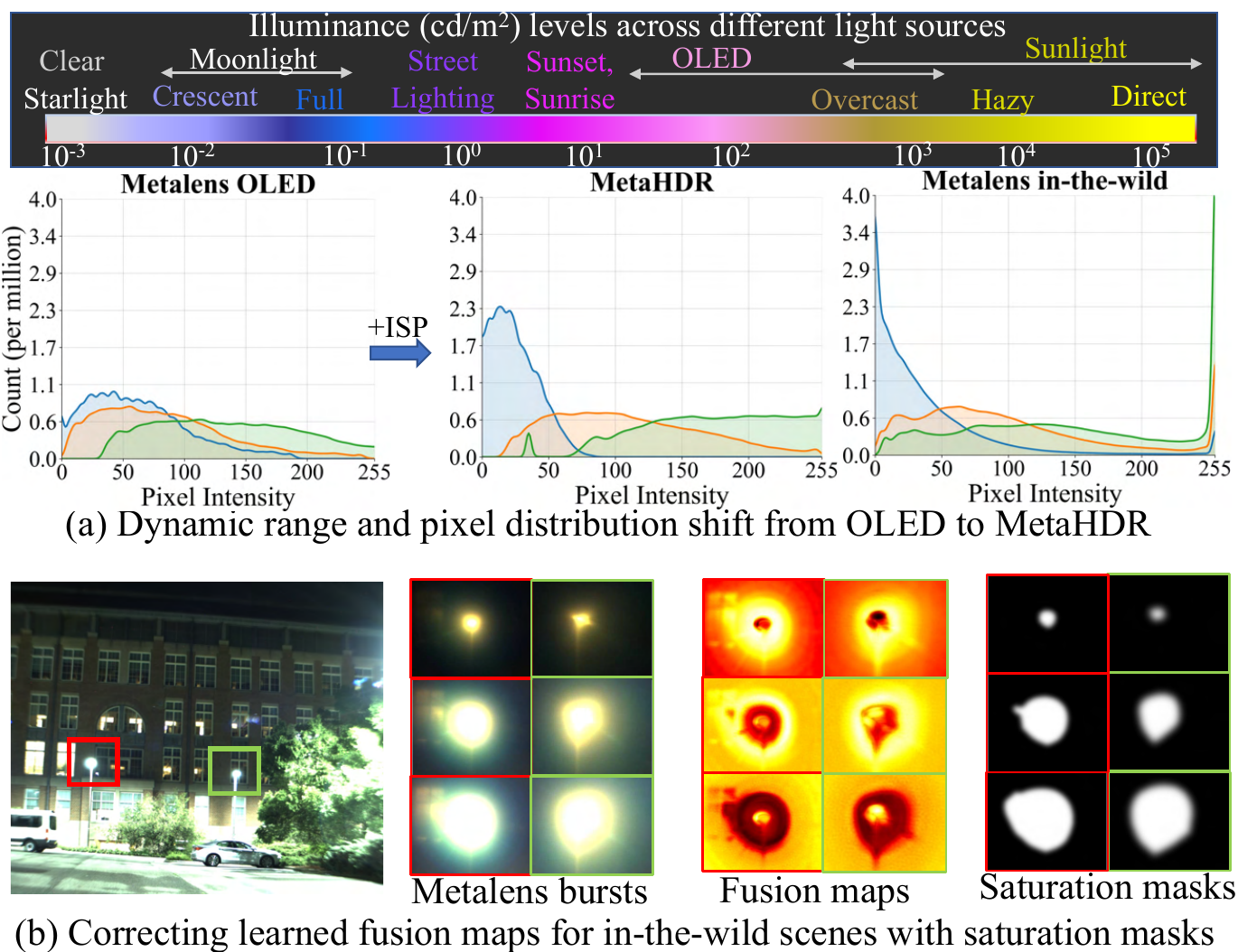}
    \caption{\textit{Fusion Maps.} 
    Switching from controlled OLED displays to real-world scenes drastically shifts luminance and pixel distributions:
    % We highlight the challenge of generalizing from OLED to in-the-wild conditions. Both luminance scales and pixel distribution varies 
    \textbf{(a)} comparison of histograms reveals this distribution gap; 
    % \textbf{(a)} when moving from indoor OLED to in-the-wild HDR scenes.
    \textbf{(b)} by fine-tuning with exposure saturation masks, we adapt fusion weights trained on OLED captures to real-world bursts.
    % \textbf{(b)} Our in-the-wild adaptation process uses saturation masks as in \cref{eq:resi_loss} to finetune the fusion maps from the weighted fusion module for real scenes.
    }
    \label{fig:itw_adapt}
    \vspace{-4mm}
\end{figure}

Collecting large-scale, paired, in-the-wild data is challenging, so prior methods often train on simulated and/or indoor OLED display captures \cite{seo2024deeplearningdrivenendtoendmetalensimaging, tseng2021neural}, resulting in overfitting and poor generalization to outdoor scenes (\cref{fig:itw_adapt}(a)).
While self-supervised strategies like BracketIRE \cite{zhang2024exposure} use pseudo-targets to adapt, they can retain implicit biases that limit generalizability. 

% As collecting large-scale paired data for in-the-wild images is challenging, existing methods typically rely on training with simulated data captured under controlled indoor OLED setups. 
% However, training on simulated images often leads to overfitting and poor generalization to in-the-wild data. BracketIRE \cite{zhang2024exposure} addresses this via self-supervised adaptation using pseudo-targets created from the network, but implicit biases can still remain and limit generalizability. Recent metalens-based methods \cite{seo2024deeplearningdrivenendtoendmetalensimaging, tseng2021neural} rely on captured data using an indoor OLED setup for training, which generalize poorly to outdoor scenes due to distribution shifts (\cref{fig:itw_adapt}(a)).

To overcome these limitations, we augment our training by simulating outdoor conditions: applying randomized gain and white-balance transforms through the ISP. 
We then fine-tune only the burst fusion module in an unsupervised fashion, using the saturation-guided loss $\mathcal{L}_\text{sat}$ from \cref{eq:resi_loss}, on a small set of unpaired real captures (\cref{fig:itw_adapt}(b)). 
Our decoupled architecture allows this targeted adaptation (expanding the dynamic range from OLED displays to in-the-wild settings), correcting fusion weights in overexposed areas via pixel-wise saturation masks $S^i$ obtained via thresholding and morphological closing. 
Moreover, the feature alignment module's multi-scale design selectively recovers features across a broad luminance range, from underexposed shadows to saturated highlights (validated in \cref{sec:exp}), enabling reliable performance across diverse, in-the-wild lighting conditions and and robust generalization to real-world capture settings.

\section{Experiments}
\label{sec:exp}

\subsection{Dataset and Implementation} 
\noindent
\textbf{Training data:}
For collecting training data, we project HDR content on a wide-angle HiSense 4K OLED television (60\% brightness) in a dark room and capture multi-frame bursts.
We leverage public burst datasets (HDM-HDR \cite{froehlich2014creating}, Zurich Raw RGB \cite{ignatov2020replacing}, and Burst HDR+ \cite{hasinoff2016burst}) alongside high-resolution still-image datasets (Flickr2K, Div2K \cite{Agustsson_2017_CVPR_Workshops}). 
To better emulate real-world dynamic range, we capture metalens frames at varied exposure levels rather than post hoc gain adjustments, and generate synthetic handheld bursts via ISP inversion following Brooks et al. \cite{brooks2019unprocessing}. 
All raw bursts are stored in 12-bit Bayer RGGB format.

% \paragraph{OLED dataset} Our OLED data capture setup involves projecting images from a wide angled, high-resolution, HDR capable OLED TV set at 60\% brightness, placed inside a dark room.
% We leverage existing multi-frame burst datasets such as HDM-HDR \cite{froehlich2014creating}, Zurich Raw RGB \cite{ignatov2020replacing}, and Burst HDR+ \cite{hasinoff2016burst} for scenes with high dynamic range and non-burst datasets such as Div2K \cite{Agustsson_2017_CVPR_Workshops}, Flickr2K, featuring  scenes with high resolution content. 
% For single image datasets we capture frames with the metalens at varying exposure levels, instead of digitally modifying image gain afterwards. The higher exposure levels simulate the dynamic range of real-world conditions more closely and improve generalization to in-the-wild metalens images. Finally, to simulate camera burst handshake, we follow the ISP inversion pipeline from \cite{brooks2019unprocessing}. Raw RGB bursts are stored in 12-bit Bayer RGGB format. 

\noindent
\textbf{In-the-Wild Captures:}
We mount our metalens module and an Alvium Allied Vision 1800 U-510 machine vision camera in parallel on a Jetson Nano Orin, powered by a 5V portable power supply. 
For each scene, we run the ISP's auto-exposure to set a base exposure based on scene brightness, then capture 3-20 frame bursts with widened exposure brackets to cover high dynamic range. 
We adaptively change the gap between shortest and longest exposure times based on varying illumination conditions, ranging from street lamps at night to mixed indoor lighting. 

% \noindent\textbf{Real captures.} For our real in-the-wild dataset, we pair our metalens camera with a refractive plano-convex lens camera, mounting them in parallel on top of a Jetson Nano Orin powered by a 5V portable power source. 
% We collect bursts with size ranging from 5–20 frames for scenes with uniform to varying illumination conditions, such as street lamps at night.
% We run the camera module ISP's auto-exposure algorithm to first determine the optimal exposure time based on scene brightness. Then, using this value as a reference, we execute a burst capture script. To accommodate scenes with high dynamic range, we further increase the gap between the shortest and longest exposure times accordingly.

\noindent
\textbf{Implementation Details:} 
% Training was carried out with $\tau_1 = \tau_2 = 0.5$, a batch size of 8, and AdamW optimizer \cite{loshchilov2017decoupled} with a learning rate of $1\mathrm{e}{-4}$ and a weight decay of $1\mathrm{e}{-5}$. We train for 300 epochs on an RTX 3090 (24GB) for 2 days until convergence. 
% We implement our pipeline using PyTorch and train the model with $\tau_1 = \tau_2 = 0.5$, batch size 8, and AdamW \cite{loshchilov2017decoupled} (learning rate $1\mathrm{e}{-4}$, weight decay $1\mathrm{e}{-5}$) for 300 epochs on an RTX 3090, converging in $~$2 days.
% 
Our pipeline is implemented in PyTorch and trained for 300 epochs on an RTX 3090 (24 GB) using AdamW optimizer \cite{loshchilov2017decoupled} (learning rate $1\mathrm{e}{-4}$, weight decay $1\mathrm{e}{-5}$), batch size 8, and loss weights $\tau_1 = \tau_2 = 0.5$ Total training time is approximately two days.

% Existing multi-frame HDR datasets, such as HDM-HDR \cite{froehlich2014creating}, Zurich Raw RGB \cite{ignatov2020replacing}, and Burst HDR+ \cite{hasinoff2016burst}, feature HDR scenes and apply digital gain to create LDR equivalents. However, this differs from real-world HDR imaging, where LDR frames are captured at varying exposure settings. To bridge this gap, we introduce a large-scale HDR metalens dataset. We project images from previous HDR datasets on an OLED screen and capture frames with the metalens at varying exposure times ($\pm 1$ and $\pm 2$ EV stops), instead of digitally modifying gain. The higher EV stop simulates real-world HDR bursts more closely and helps generalize better to in-the-wild images. Finally, to simulate camera burst handshake, we follow the ISP inversion pipeline from \cite{brooks2019unprocessing}. The raw RGB bursts are stored in 12-bit Bayer RGGB format.

% \subsection{Comparison with existing works}

\begin{figure*}[!t]
    \centering
    \includegraphics[width=\linewidth]{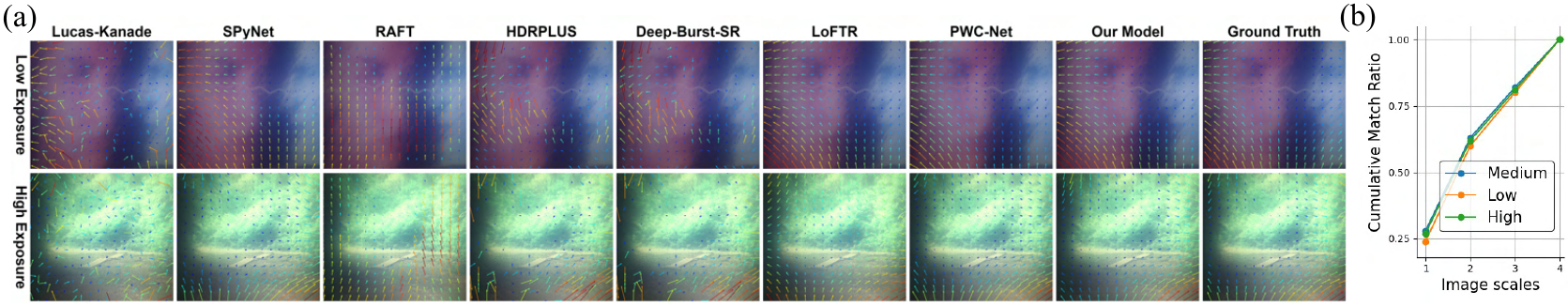}
    % \vspace{-4mm}
    \caption{\textit{Visualizing Displacement Vectors.} \textbf{(a)} Motion flow fields estimated by different burst alignment methods across varying exposure levels (zoom in for detail). \textbf{(b)} Cumulative counts of successful feature matches per pyramid level under different exposures, highlighting our multi-scale enhancement's robustness.}
    \label{fig:Align_arrows}
    \vspace{-7mm}
\end{figure*}

\subsection{Benchmarking Against State-of-the-Art}  
\begin{table}[t]
\scriptsize
\centering
\setlength{\tabcolsep}{1pt}
\renewcommand{\arraystretch}{0.85}
\caption{\textit{Benchmark on OLED Dataset.} Quantitative comparison across different restoration method categories.}
\label{tab:comparison}
\begin{tabular}{l|c|cccc|ccc}
\toprule
\multirow{2}{*}{Methods} & \multirow{2}{*}{Year} & \multicolumn{4}{c|}{\dataName{} } & \multirow{2}{*}{\makecell{Params\\(M)}} & \multirow{2}{*}{\makecell{MACs\\(G)}} & \multirow{2}{*}{\makecell{Time\\(s)}} \\
\cmidrule{3-6}
 & & PSNR$\uparrow$ & SSIM$\uparrow$ & LPIPS$\downarrow$ & NIQE$\downarrow$ & & & \\
\midrule
$\circ$ ADNet~\cite{TIAN2020117} & 2020 & 23.24 & 0.73 & 0.36 & 5.60 & 3.81 & 212.3 & \textbf{0.06} \\
% & HDRUNet~\cite{chen2021hdrunetsingleimagehdr} & 2021 & 18.37 & 0.62 & 0.46 & 6.09 & -- & -- & -- & -- & -- & -- & -- \\
$\circ$ AHDRNet~\cite{yan2019attention} & 2019 & 22.37 & 0.72 & 0.37 & 5.47 & 2.73 & 312.5 & \textbf{0.06} \\
$\circ$ SCTNet~\cite{tel2023alignment} & 2023 & 24.21 & 0.74 & 0.32 & 5.45 & \textbf{0.95} & 144.7 & 0.94 \\
$\circ$ HDR-Tran~\cite{liu2022ghost} & 2022 & 24.58 & 0.74 & 0.33 & 5.45 & \underline{1.19} & 352.2 & 0.65 \\
$\circ$ BracketIRE~\cite{zhang2024exposure} & 2024 & 23.46 & 0.73 & 0.33 & 5.31 & 9.51 & 1813 & 0.13 \\
% \midrule
% \multirow{2}{*}{\makecell[l]{Plug-and-\\Play}} & DPIR~\cite{zhang2021plugandplayimagerestorationdeep} & 2021 & -- & -- & -- & -- & -- & -- & -- & -- & -- & -- & -- \\
% & GSPnP & -- & -- & -- & -- & -- & -- & -- & -- & -- & -- & -- \\
% & DiffPIR & -- & -- & -- & -- & -- & -- & -- & -- & -- & -- & -- \\
% \midrule
\midrule
$\bullet$ BSRT \cite{luo2022bsrt} & 2022 & 23.41 & 0.73 & 0.38 & 5.74 & 19.62 & 3433 & 0.77 \\
$\bullet$ EBSR \cite{luo2021ebsr} & 2021 & 18.34 & 0.59 & 0.50 & 5.36 & 23.67 & 4242 & 0.29 \\
$\bullet$ DBSR~\cite{bhat2021deepburstsuperresolution} & 2021 & 22.44 & 0.71 & 0.37 & \underline{5.29} & 12.88 & 1386 & 0.10 \\
$\bullet$ HDR-USRNet ~\cite{lecouat2022high} & 2020 & 23.25 & 0.69 & 0.41 & 6.51 & 24.96 & 768.9 & 0.09 \\
$\bullet$ HCDeblur~\cite{rim2024deephybridcameradeblurring} & 2024 & 24.12 & 0.73 & 0.32 & 5.53 & 11.61 & 69.11 & 0.09 \\
\midrule
$\star$ SwinIR \cite{liang2021swinir} & 2021 & 22.10 & 0.70 & 0.43 & \textbf{5.17} & 11.47 & 1693 & 1.06 \\
$\star$ NAFNet~\cite{chen2022simplebaselinesimagerestoration} & 2022 & \underline{25.07} & \underline{0.75} & \underline{0.30} & 5.82 & 176.5 & 275.5 & 0.08 \\
$\star$ Restormer \cite{zamir2022restormer} & 2022 & 25.38 & \underline{0.75} & 0.31 & 5.85 & 26.1 & 317.6 & 0.19 \\
$\star$ ESRGAN \cite{wang2021real} & 2021 & 22.34 & 0.72 & 0.35 & 5.36 & 26.63 & 3925 & 0.3 \\
\midrule
$\triangle$ MultiWiener-Net \cite{yanny2022deep} & 2022 & 21.31 & 0.67 & 0.42 & 5.90 & 6.72 & 89.57 & 0.06 \\
% % & NeuralNano~\cite{tseng2021neural} & 2021 & -- & -- & -- & -- & -- & -- & -- & -- & -- & -- & -- \\
$\triangle$ EIDL-DRMI~\cite{seo2024deeplearningdrivenendtoendmetalensimaging} & 2024 & 21.45 & 0.68 & 0.31 & 5.59 & 58.10 & 108.4 & 0.09 \\
$\triangle$ NNOptic~\cite{tseng2021neural} & 2021 & 19.4 & 0.61 & 0.48 & 5.76 & 29.10 & \textbf{12.87} & 0.10 \\
\midrule
\rowcolor{green!20}\textbf{Ours} & 2025 & \textbf{27.52} & \textbf{0.81} & \textbf{0.23} & 5.43 & 12.3 & \underline{66.42} & \underline{0.08} \\
\bottomrule
\end{tabular}
\begin{tablenotes}[flushleft] 
\scriptsize
\item[1] $\circ$: HDR Fusion, $\bullet$: Non-HDR Burst Fusion, $\star$: General Restoration, $\triangle$: Metalens.
\end{tablenotes}
\vspace{-3mm}
\end{table}

We benchmark our burst fusion against leading burst fusion, HDR fusion, general restoration, and metalens-specific methods using PSNR, SSIM \cite{ssim}, and LPIPS \cite{lpips}. 
We use reference-free metrics (NIQE \cite{niqe}, BRISQUE \cite{brisque}, and PIQE \cite{venkatanath2015blind}) for unpaired real-world data.

% We compare our burst fusion algorithm with prior burst fusion methods, HDR fusion approaches, general image restoration frameworks, and metalens-specific restoration algorithms. For evaluation, we utilize metrics including PSNR, SSIM \cite{ssim}, and LPIPS \cite{lpips}, and non-reference metrics like NIQE \cite{niqe}, BRISQUE \cite{brisque}, and PIQE \cite{venkatanath2015blind} for real-world scenarios lacking reliable ground truth.

As shown in \cref{tab:comparison}, HDR fusion methods improve over non-HDR burst methods on our OLED-metalens dataset but still fall short of general purpose restoration methods like Restormer \cite{zamir2022restormer} and NAFNet \cite{chen2022simplebaselinesimagerestoration}.
Surprisingly, metalens-tailored algorithms also underperform, likely due to their reliance on fixed illumination conditions.
In contrast, our joint burst fusion and restoration pipeline consistently achieves superior performance across most metrics while maintaining low runtime and computational cost (see \cref{fig:teaser}(c)).
Visual comparisons are provided in the Supplementary Material, further validating our framework's effectiveness in balancing quality and efficiency.

% The quantitative results and runtime comparisons are presented in \cref{tab:comparison}. To better illustrate the trade-off between performance and efficiency, we visualize performance versus runtime and computational cost in \cref{fig:teaser}(c). HDR fusion methods outperform most non-HDR burst fusion approaches on our OLED metalens dataset but remain $\sim1$ dB lower in PSNR compared to general restoration methods such as Restormer \cite{zamir2022restormer} and NAFNet \cite{chen2022simplebaselinesimagerestoration}. 
% Surprisingly, we find none of the approaches proposed for metalens restoration to reach good performance. This could be attributed to the fact that these methods were not designed to handle varying illumination levels as featured in our ISP augmented metalens OLED dataset.
% In contrast, our method consistently achieves superior performance across most evaluation metrics while maintaining computational efficiency. 
% Visual results are provided in the supplementary material. These findings validate the effectiveness of our joint burst fusion and restoration framework.

% \subsection{Evaluation of exposure mask quality} 
\subsection{Exposure Mask Quality Assessment}
\begin{figure}[t!]
    \centering
    \includegraphics[width=0.99\linewidth]{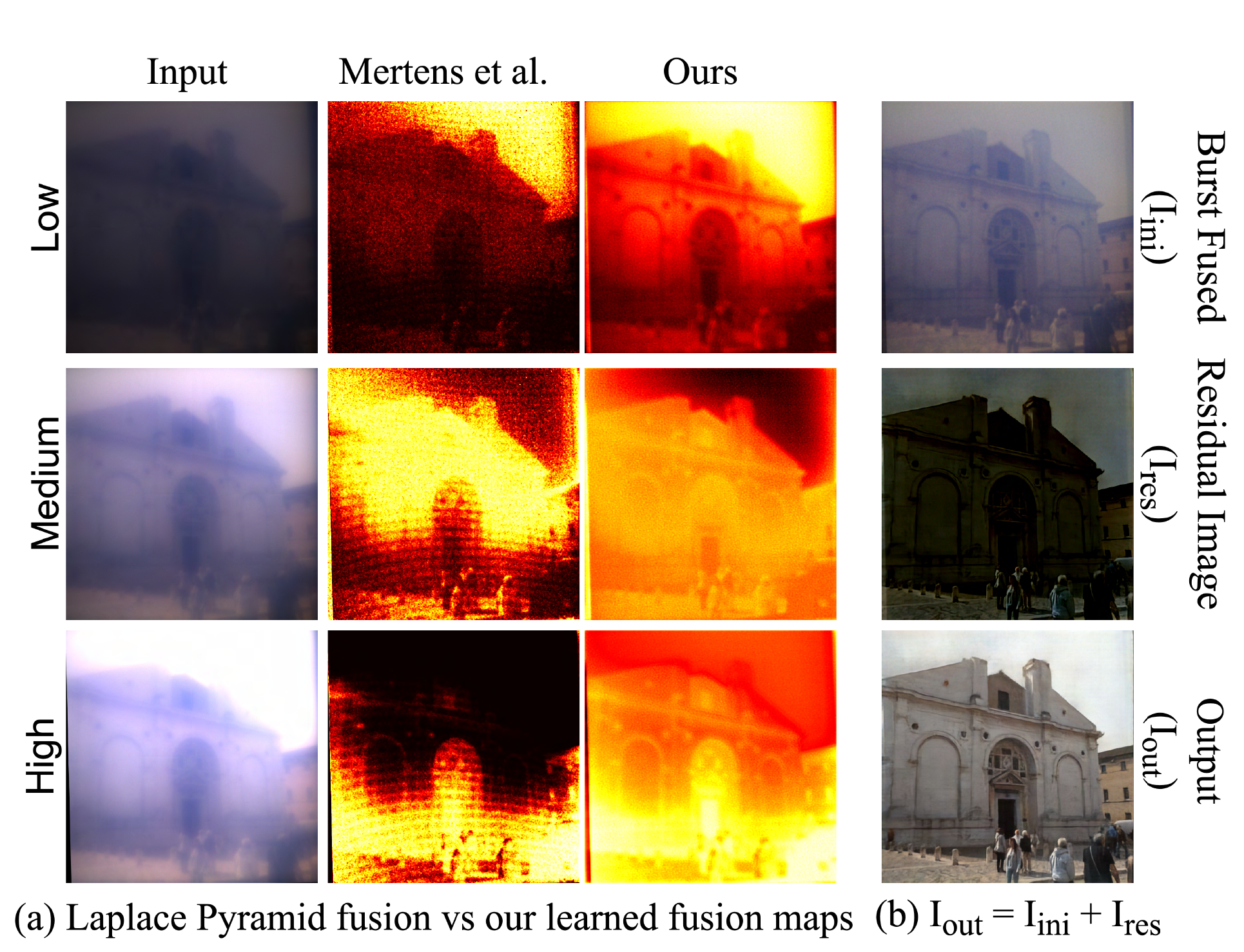}
    % \vspace{-4mm}
    \caption{\textit{Model Analysis.} 
    \textbf{(a)} Traditional Laplacian pyramid fusion \cite{mertens2007exposure} yields noisy, artifact-prone exposure fusion weight maps, whereas our learned fusion weights are smooth and well-structured with significantly less noise and artifacts. 
    \textbf{(b)} The fusion maps produce an initial image estimate $I_\text{init}$, which is subsequently refined by the residual $I_\text{res}$ to generate the final output image $I_\text{out}$.}
    \vspace{-3mm}
    \label{fig:expmap}
\end{figure}
We evaluate our learned exposure fusion maps $w^i$ (\cref{eq:residual}) against classic intensity-based weights from Mertens et al. \cite{mertens2007exposure}. 
As shown in \cref{fig:expmap}(a), relying solely on pixel intensity produces noisy textures in exposure maps, and ringing artifacts, particularly around blurred regions (see \cref{fig:cam}(b)), all of which our learned weights successfully avoid. 
This improvement stems from end-to-end training on metalens multi-exposure bursts, enabling the network to assign smooth, content-adaptive weights.
More visual comparisons are provided in Supplementary Material.

% To assess the quality of our learned exposure fusion maps $w^i$  (used in \cref{eq:residual}) we compare it against exposure maps generated from Mertens et al. \cite{mertens2007exposure}, where simply relying on pixel intensity as a measure for fusion weight can lead to formation of noisy textures in the exposure map as shown in \cref{fig:expmap} (a). Further, the blur in the image leads to ring-like artifacts similar to what we observe in \cref{fig:cam}(b), which do not appear in our learned exposure weights. This is due to the learned weights becoming increasingly adapted to metalens multi-exposure images during training. More visualizations are given in the supplementary.

\begin{table}[!htbp]
% \normalsize
\centering
\renewcommand{\arraystretch}{0.65} 
\setlength{\tabcolsep}{5pt}
\setlength{\extrarowheight}{5pt}
\scriptsize
\caption{\textit{Comparison of Burst Alignment Methods.} Each metric is shown with its mean $\pm$ standard deviation.}
\begin{tabular}{l|cccc}
\toprule
Method & \makecell{Mean$\downarrow$} & \makecell{Cosine$\uparrow$} & \makecell{Median$\downarrow$} & \makecell{CPU Time(s)$\downarrow$} \\
\midrule
RAFT~\cite{teed2020raftrecurrentallpairsfield} & 47.78$\pm$56.47 & 0.68$\pm$0.55 & 54.25$\pm$93.67 & 14.15 \\
Lucas-Kanade~\cite{10.5555/1623264.1623280} & 15.86$\pm$5.40 & 0.41$\pm$0.64 & 22.89$\pm$8.27  & \textbf{0.02} \\
HDRPLUS~\cite{10.1145/2980179.2980254} & 8.77$\pm$3.66 & 0.76$\pm$0.45 & 9.11$\pm$6.90 & 25.03 \\
Deep-Burst-SR~\cite{ipol.2023.460} & 8.76$\pm$3.66 & 0.76$\pm$0.46 & 9.10$\pm$6.90 & 1.60  \\
LoFTR~\cite{sun2021loftrdetectorfreelocalfeature} & 2.83$\pm$2.27 & 0.98$\pm$0.09 & 4.29$\pm$3.47 & 13.80 \\
SPyNet~\cite{ranjan2016opticalflowestimationusing} & 2.38$\pm$1.80 & 0.97$\pm$0.16 & 2.00$\pm$1.70 & 2.70 \\
PWC-Net~\cite{sun2018pwcnetcnnsopticalflow} & 1.66$\pm$2.64 & 0.98$\pm$0.13 & \textbf{0.93}$\pm$\textbf{0.49} & 0.17  \\
\rowcolor{green!20} \textbf{Ours}& \textbf{1.34}$\pm$\textbf{0.89} & \textbf{0.99}$\pm$\textbf{0.06} & 1.92$\pm$1.19 & 0.11 \\
\bottomrule
\end{tabular}
\label{tab:Align_Comparison}
\vspace{-3mm}
\end{table}

\subsection{Burst Alignment Evaluation} 
\begin{figure}[t]
    \centering
    \includegraphics[width=\linewidth]{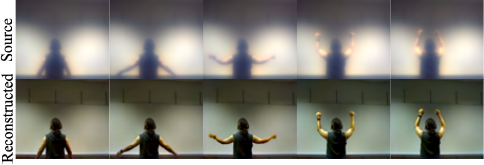}
    % \caption{\textbf{(a)} Luminance range across 
    \caption{\textit{Real-time Video Restoration.} (Top row) Raw burst frames captured at the reference exposure.
    (Bottom row) Corresponding reconstructed frames output by our pipeline in real time, demonstrating high-quality restoration under dynamic motion.
    % \todo{Add labels to images or describe what the images in the first and second rows are.}
    }
    \vspace{-2mm}
    \label{fig:vid}
\end{figure}

Accurate alignment is critical for high-quality multi-frame image restoration. 
We benchmark our alignment module on the OLED dataset against standard algorithms, following Brooks et al. \cite{brooks2019unprocessing}, and report results in \cref{tab:Align_Comparison}. 
Our method achieves lowest registration errors and exhibits minimal run-to-run variance, demonstrating both accuracy and stable performance. 
\Cref{fig:Align_arrows}(a) shows that, unlike HDR+ \cite{hasinoff2016burst}, which assumes constant exposure and struggles under varying exposures, our alignment algorithm produces more accurate flow fields and outperforms other baselines at both low and high exposure levels. 
Further, \cref{fig:vid} demonstrates the effectiveness of our alignment method in real-world video sequence with a moving subject. 
Additional results are in the Supplementary Material.

% As the final restoration quality is highly sensitive to the alignment procedure, we evaluate the effectiveness our alignment module. Due to the lack of real paired metalens burst datasets, we follow \cite{brooks2019unprocessing} and evaluate our burst alignment algorithm against others on our OLED dataset. 
% As shown in \cref{tab:Align_Comparison}, our method outperforms existing image alignment algorithms across most distance metrics. We further evaluate run-to-run variations and observe that our burst alignment algorithm yields consistently low deviations, demonstrating robust and stable performance. 
% \Cref{fig:Align_arrows}(a) visualizes the estimated flow fields from different approaches. Notably, HDR+ \cite{hasinoff2016burst} assumes a constant exposure level, which limits its effectiveness under varying exposures. We evaluate our alignment algorithm across low and high exposure levels and retain superior margins against prior methods, with detailed results provided in the supplementary. 
% Finally, \cref{fig:vid} shows the effectiveness of our alignment method in real video capture scenarios at with a moving human subject.

% \todo{What does this analysis convey?} \todo{It shows xxx...} \todo{\textit{moved that sentence to 4.1 second para}}

\setlength{\tabcolsep}{10pt}
\begin{table}[t]
\centering
\scriptsize
\caption{\textit{Benchmark on Real Data.} Quantitative evaluation of best methods on real in-the-wild scenes using reference-free metrics.}
\label{tab:hdr_evaluation}
% \resizebox{\columnwidth}{!}{%
\def\arraystretch{1.05}%
\begin{tabular}{lccc}
\toprule
% \multirow{2}{*}{\textbf{Method}} & \multicolumn{4}{c}{\textbf{Non-Reference Metrics}} \\
% \cmidrule{2-5}
Method & BRISQUE $\downarrow$ & NIQE $\downarrow$ & PIQE $\downarrow$  \\
\midrule
HCDeblur \cite{rim2024deephybridcameradeblurring} & 42.59 & 11.99 & 59.65 \\
Restormer \cite{zamir2022restormer} & 28.85 & 6.06 & 40.18  \\
HDR-Tran \cite{liu2022ghost} & 47.64 &  7.28 & 48.90  \\
NAFNet \cite{chen2022simplebaselinesimagerestoration} & 29.17 & 5.88 & 37.53 \\
% HDR-USRNet \cite{} & -- & -- & -- \\

\midrule
\rowcolor{green!20} \textbf{Ours} & \textbf{23.75} & \textbf{4.31} & \textbf{28.77} \\
\bottomrule
\end{tabular}%
\vspace{-2mm}
% }
\end{table}
\begin{figure*}[!t]
    \centering
    \includegraphics[width=.95\linewidth]{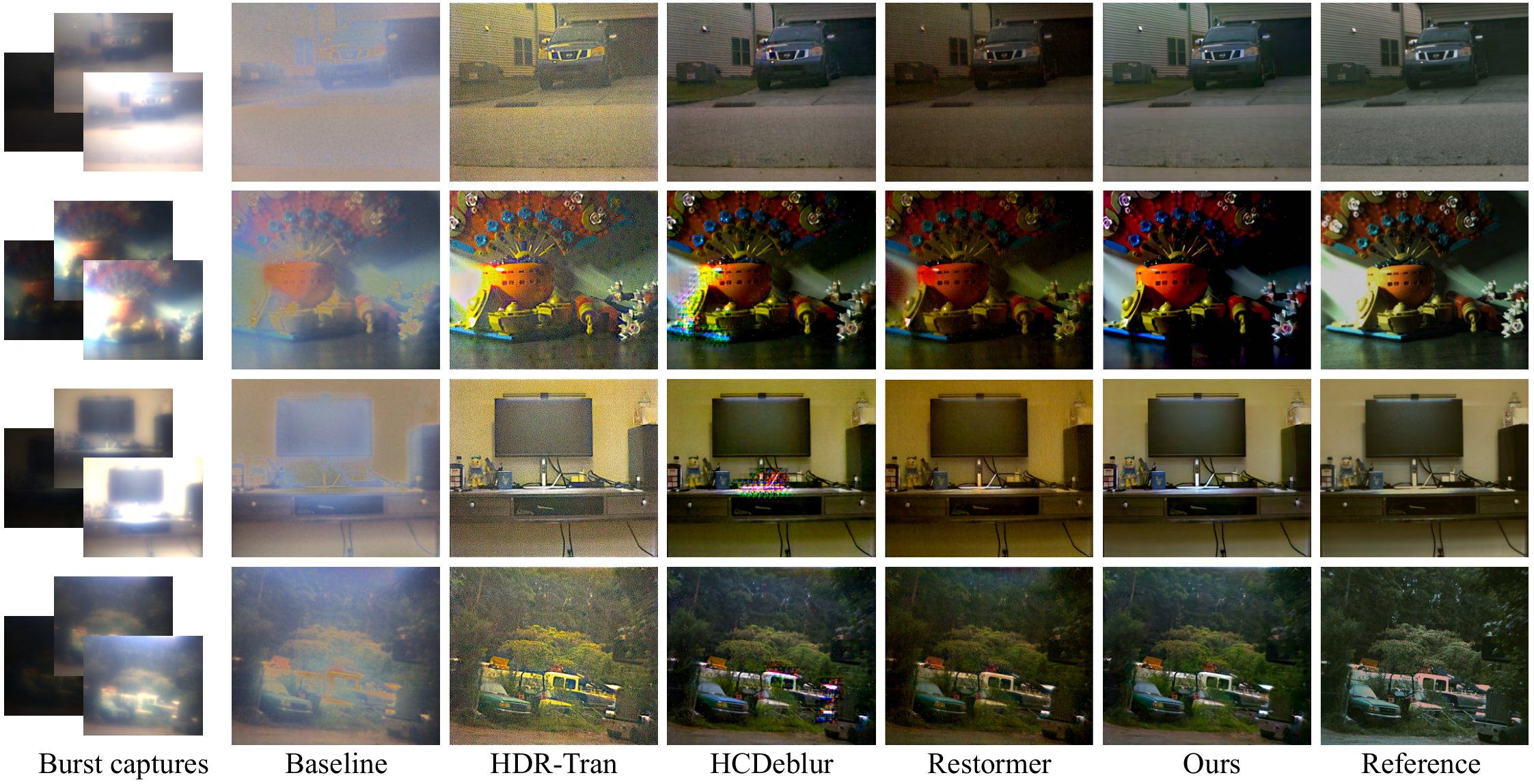}
    % \vspace{-4mm}
    \caption{\textit{In-the-Wild Results.} 
    We show real-world captures from our handheld metalens camera prototype. 
    % Performance on real captured scenes with handheld prototype.
    Our baseline restoration uses Wiener deconvolution with fusion from Mertens et al. \cite{mertens2007exposure}. Reference images are from a conventional compound-lens camera.
    }
    \vspace{-3mm}
    \label{fig:itw_main}
\end{figure*}

 \begin{figure}[t]
    \centering
    \includegraphics[width=\linewidth]{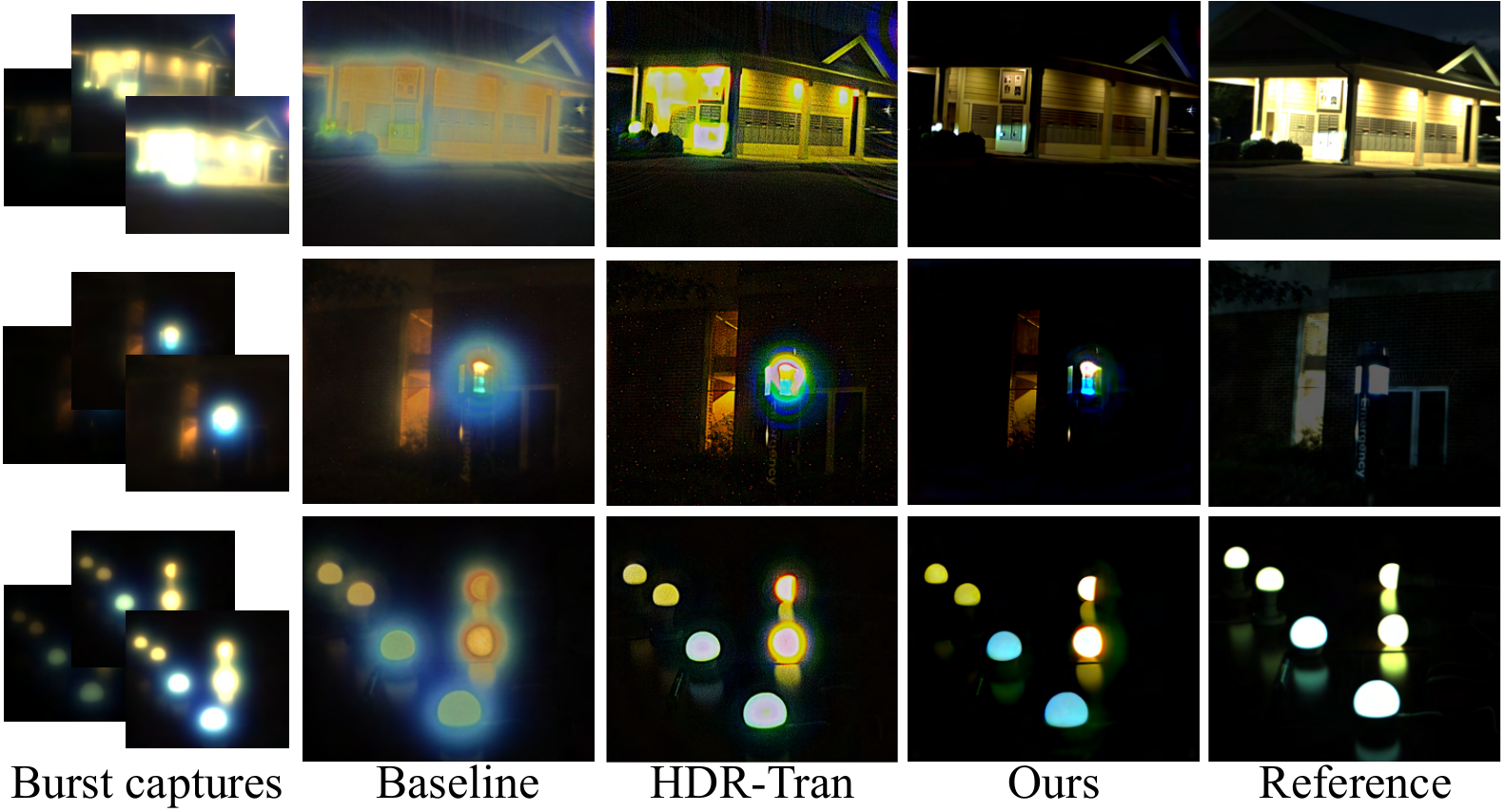}
    % \includesvg[width=\linewidth]{figures/assets/hdr_main2.svg}
    \caption{\textit{HDR Restoration.}
    Qualitative results on in-the-wild HDR scenes, showcasing our method's ability to recover both dark shadows and bright highlights in challenging environments.}
    \label{fig:itw_hdr}
    \vspace{-5mm}
\end{figure}

\subsection{In-the-Wild Zero-Shot Generalization}

To assess real-world performance, we apply the best performing models from \cref{tab:comparison} directly to unpaired, in-the-wild bursts and evaluate their zero-shot generalization performance using no-reference metrics in \cref{tab:hdr_evaluation} and visualize qualitative comparisons in \cref{fig:itw_main}.
Our approach produces noticeably sharper, aberration-free images across diverse lighting, outperforming all baselines on every metric. 
Moreover, as shown in \cref{fig:itw_hdr}, our method successfully handles extreme HDR scenes of dark and bright regions. This robust generalization stems from our multi-scale feature alignment and unsupervised fusion adaptation (\cref{sec:generalize_itw}). 
See Supplementary Material for more examples.

% While \cref{tab:comparison} evaluates the performance of models in a supervised setting, the synthetic evaluations do not reflect in-the-wild performance. 
% In addition to evaluations performed on synthetic dataset, we further conduct evaluations on real in-the-wild images.
% Specifically, we test top-performing models from \cref{tab:comparison} and evaluate their zero-shot generalization to in-the-wild scenes. Due to the lack of ground truth data, we utilize non-reference metrics to assess the quality of the restored image. Quantitative and qualitative results are reported in \cref{tab:hdr_evaluation} and \cref{fig:itw_main} respectively. 
% As shown, our results look sharper and aberration-free and can work across arbitrary scene brightness as presented in Fig. 3 in the supplementary. 
% Supported by our efforts in \cref{sec:generalize_itw} to augment the dataset with varying white balance factors, we notice that our method maintains better color fidelity to the reference ground truth in \cref{fig:itw_main}. 
% Also, since our framework possesses both HDR and aberration removal characteristics it performs better against all top performers from \cref{tab:comparison} . We also present visual results on HDR scenes featuring dark and bright regions in \cref{fig:itw_hdr}. We compare against an HDR fusion method HDR-Tran \cite{liu2022ghost} which fails to accurately reconstruct regions around very bright regions and doesn't remove noise from the image either. More results are provided in the supplementary.

 \begin{figure}[t]
    \centering
    \vspace{-2mm}
    \includegraphics[width=\linewidth]{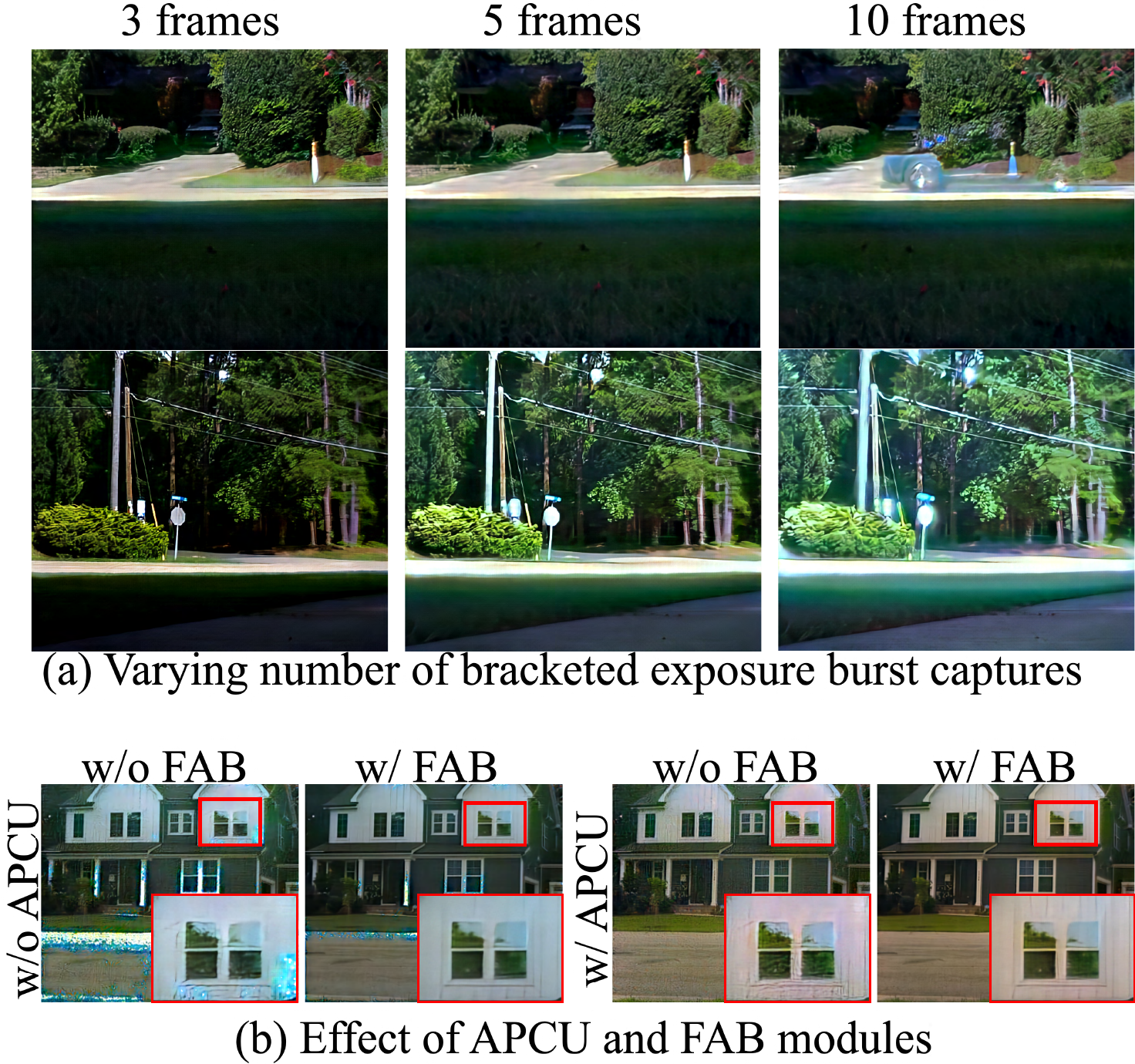}
    % \caption{\textbf{(a)} Luminance range across 
    \caption{\textit{Ablation Study}. 
    \textbf{(a)} Burst size trade-off: too few frames yield underexposed results, while too many introduce motion blur and ghosting from fast-moving objects.
    % In real scenes, lower burst sizes lead to darker outputs while more frames can lead to motion blur or ghosting artifacts due to fast moving objects. 
    \textbf{(b)} Removing the Adaptive Pixel Correction Unit (left) or Feature Alignment Block (right) degrades restoration quality, leaving visible artifacts.
    % Effect of removing the adaptive pixel correction unit (APCU) (left) and feature alignment block (FAB) (right).
    }
    \label{fig:abla}
    \vspace{-5mm}
\end{figure}

\setlength{\tabcolsep}{10pt}
\begin{table}[t]
\centering
\scriptsize
\caption{\textit{Ablation study on model design.} For each experiment, the base starting model was obtained from a single training run of \cref{tab:comparison}. Base model number \underline{underlined}.}
\label{tab:ablation}
% \resizebox{\columnwidth}{!}{%
\def\arraystretch{1.05}%
\begin{tabular}{lcc}
\toprule
% \multirow{2}{*}{\textbf{Method}} & \multicolumn{4}{c}{\textbf{Non-Reference Metrics}} \\
% \cmidrule{2-5}
Experiments   & PSNR $\uparrow$ & LPIPS $\downarrow$  \\
\midrule
Number of burst frames (3 \textbar 5 \textbar 10)  & \underline
{26.5} \textbar 26.9 \textbar 26.7 & \underline{0.25} \textbar 0.24 \textbar 0.24 \\
Channel depth (16 \textbar  20 \textbar  24) & 24.3 \textbar \underline{26.5} \textbar 26.3 & 0.29 \textbar \underline{0.25} \textbar 0.25  \\
Transformers per AFB (1 \textbar  2 \textbar  3) & 25.8 \textbar \underline{26.5} \textbar 26.5 & 0.28 \textbar \underline{0.25} \textbar 0.26  \\
w/o. (APCU \textbar  AFB) & 24.7 \textbar 24.4 & 0.38 \textbar 0.40 \\
w/o. APCU \& w/o. AFB  & 24.2 & 0.41 \\
% HDR-USRNet \cite{} & -- & -- & -- \\
w/o. $I_{res}$  & 19.9 & 0.47 \\
% \midrule
% \rowcolor{green!20} \textbf{Ours} & \textbf{23.75} & \textbf{4.31} & \textbf{28.77} \\
\bottomrule
\end{tabular}%
\vspace{-6mm}
% }
\label{tab:ablat}
\end{table}

\subsection{Ablation studies}

We analyze key components through ablation studies shown in \cref{fig:abla} and \cref{tab:ablation}. \Cref{fig:abla}(a) demonstrates optimal performance at 5 burst frames (26.9 dB PSNR), with additional frames degrading results due to burst motion artifacts. \Cref{fig:abla}(b) shows that removing the Adaptive Pixel Correction Unit (APCU) allows corrupted pixels to propagate, reducing performance to 24.7 dB. Similarly, disabling Feature Alignment Module (FAB) introduces
artifacts in the recovered image from incorrectly estimated
residuals. The restoration module ($I_{\text{res}}$) proves essential, with its removal causing dramatic degradation to 19.9 dB PSNR. Several additional ablations are given in Supplementary Material.

\section{Conclusions}
\vspace{-3mm}
We have shown that a single ultra-thin nanophotonic metalens combined with an efficient burst-fusion and restoration pipeline can achieve image quality on par with conventional multi-element lenses, even in challenging in-the-wild handheld scenarios. 
Extensive evaluations on both controlled and real-world datasets demonstrate that our approach not only outperforms state-of-the-art restoration methods but also runs in real time on edge hardware.
We believe that our work paves the way for truly miniaturized, high-performance imaging in next-generation AR/VR, wearable, and IoT applications.

% We believe our work demonstrates the feasibility of practical, real-world imaging with ultra-thin metalens optics combined with lightweight burst-based computation, establishing a path toward deployable metasurface cameras in next-generation consumer devices.

%%%%%%%%% REFERENCES
{\small
\bibliographystyle{ieee_fullname}
\bibliography{main}
}

\end{document}